\documentclass[12pt,a4paper]{article}
\usepackage{amsmath,amssymb}
\newcommand{\beq}{\begin{eqnarray}}
\newcommand{\eeq}{\end{eqnarray}}

\newcommand{\be}{\begin{equation}}
\newcommand{\ee}{\end{equation}}

\newcommand{\bm}{\begin{multline}}
\newcommand{\fm}{\end{multline}}

\begin{document}
\setlength{\unitlength}{.8mm}

\begin{titlepage} 
\vspace*{0.5cm}
\begin{center}
{\Large\bf Finite size effects in the SS-model:  two component nonlinear integral equations}
\end{center}
\vspace{2.5cm}
\begin{center}
{\large \'Arp\'ad Heged\H us}
\end{center}
\bigskip
\begin{center}
Research Institute for Particle and Nuclear Physics,\\
Hungarian Academy of Sciences,\\
H-1525 Budapest 114, P.O.B. 49, Hungary\\ 
\end{center}
\vspace{3.2cm}
\begin{abstract}
We derive a finite set of nonlinear integral equations for describing the finite size dependence of the ground state
energy of the $O(4)$ nonlinear sigma model. By modifying the kernel functions of these equations we propose nonlinear
integral equations for  the finite size effects in the SS-model. 
The equations are formulated in terms of two complex valued unknown functions and they are valid for 
arbitrary real values of the couplings. 
\end{abstract}

\end{titlepage}



\newsavebox{\SSa}
\sbox{\SSa}{\begin{picture}(140,25) (-70,-12.5)

\put(0,0){\circle*{3}}
\put(10,0){\circle{3}}
\put(20,0){\circle{3}}
\put(30,0){\circle{3}}
\put(40,10){\circle{3}}
\put(40,-10){\circle{3}}
\put(-10,0){\circle{3}}
\put(-20,0){\circle{3}}
\put(-30,0){\circle{3}}
\put(-40,10){\circle{3}}
\put(-40,-10){\circle{3}}

\put(1.5,0){\line(1,0){7}}
\put(21.5,0){\line(1,0){7}}
\put(-8.5,0){\line(1,0){7}}
\put(-28.5,0){\line(1,0){7}}

\put(31.1,1.1){\line(1,1){7.7}}
\put(31.1,-1.1){\line(1,-1){7.7}}
\put(-31.1,1.1){\line(-1,1){7.7}}
\put(-31.1,-1.1){\line(-1,-1){7.7}}

\multiput(12.5,0) (1,0) {6} {\circle*{0.2}}
\multiput(-17.5,0) (1,0) {6} {\circle*{0.2}}

\put(2,-3){\makebox(0,0)[t]{{\protect\scriptsize 0}}}
\put(12,-3){\makebox(0,0)[t]{{\protect\scriptsize 1}}}
\put(22,-3){\makebox(0,0)[t]{{\protect\scriptsize $p$--3}}}
\put(30,-3){\makebox(0,0)[t]{{\protect\scriptsize  $p$--2}}}
\put(44,-11){\makebox(0,0)[t]{{\protect\scriptsize $p$--1}}}
\put(44,11){\makebox(0,0)[t]{{\protect\scriptsize $p$}}}
\put(-8,-3){\makebox(0,0)[t]{{\protect\scriptsize --1}}}
\put(-18,-3){\makebox(0,0)[t]{{\protect\scriptsize 3--$\tilde{p}$}}}
\put(-28,-3){\makebox(0,0)[t]{{\protect\scriptsize 2--$\tilde{p}$}}}
\put(-45,-10){\makebox(0,0)[t]{{\protect\scriptsize 1--$\tilde{p}$}}}
\put(-45,11){\makebox(0,0)[t]{{\protect\scriptsize --$\tilde{p}$}}}

\end{picture}}



\newsavebox{\SSddv}
\sbox{\SSddv}{\begin{picture}(140,25) (-70,-12.5)
\setlength{\unitlength}{.5mm} 
\put(0,0){\circle*{3}}
\put(10,0){\circle{3}}
\put(20,0){\circle{3}}
\put(30,0){\circle{3}}
\put(40,10){\circle{3}}
\put(40,-10){\circle{3}}
\put(-10,0){\circle{3}}
\put(-20,0){\circle{3}}
\put(-30,0){\circle{3}}
\put(-40,10){\circle{3}}
\put(-40,-10){\circle{3}}

\put(1.5,0){\line(1,0){7}}
\put(21.5,0){\line(1,0){7}}
\put(-8.5,0){\line(1,0){7}}
\put(-28.5,0){\line(1,0){7}}

\put(31.1,1.1){\line(1,1){7.7}}
\put(31.1,-1.1){\line(1,-1){7.7}}
\put(-31.1,1.1){\line(-1,1){7.7}}
\put(-31.1,-1.1){\line(-1,-1){7.7}}

\put(58,-2){$a$}

\multiput(12.5,0) (1,0) {6} {\circle*{0.2}}
\multiput(-17.5,0) (1,0) {6} {\circle*{0.2}}
\setlength{\unitlength}{1mm} 

\thicklines{
\put(13.5,0){\oval(28,15)[]}
\put(-15.8,0){\oval(23,15)[]}
}
\setlength{\unitlength}{.5mm} 
\put(2,-3){\makebox(0,0)[t]{{\protect\tiny 0}}}
\put(12,-3){\makebox(0,0)[t]{{\protect\tiny 1}}}

\put(29,-3){\makebox(0,0)[t]{{\protect\tiny  $p$--2}}}
\put(44,-10.5){\makebox(0,0)[t]{{\protect\tiny $p$--1}}}
\put(44,11){\makebox(0,0)[t]{{\protect\tiny $p$}}}
\put(-12,-3){\makebox(0,0)[t]{{\protect\tiny --1}}}

\put(-28,-3){\makebox(0,0)[t]{{\protect\tiny 2--$\tilde{p}$}}}
\put(-45,-10){\makebox(0,0)[t]{{\protect\tiny 1--$\tilde{p}$}}}
\put(-45,11){\makebox(0,0)[t]{{\protect\tiny --$\tilde{p}$}}}
\put(28,-17){\makebox(0,0)[t]{{\protect\tiny $a_0(x),\bar{a}_0(x),G_p(x)$}}}
\put(-29,-17){\makebox(0,0)[t]{{\protect\tiny $a(x),\bar{a}(x),G_{\tilde{p}-1}(x)$}}}

\end{picture}}



\newsavebox{\SSnlie}
\sbox{\SSnlie}{\begin{picture}(140,25) (-70,-12.5)
\setlength{\unitlength}{.5mm} 
\put(0,0){\circle*{3}}
\put(10,0){\circle{3}}
\put(20,0){\circle{3}}
\put(30,0){\circle{3}}
\put(40,10){\circle{3}}
\put(40,-10){\circle{3}}
\put(-10,0){\circle{3}}
\put(-20,0){\circle{3}}
\put(-30,0){\circle{3}}
\put(-40,10){\circle{3}}
\put(-40,-10){\circle{3}}

\put(1.5,0){\line(1,0){7}}
\put(21.5,0){\line(1,0){7}}
\put(-8.5,0){\line(1,0){7}}
\put(-28.5,0){\line(1,0){7}}

\put(31.1,1.1){\line(1,1){7.7}}
\put(31.1,-1.1){\line(1,-1){7.7}}
\put(-31.1,1.1){\line(-1,1){7.7}}
\put(-31.1,-1.1){\line(-1,-1){7.7}}

\put(58,-2){$b$}

\multiput(12.5,0) (1,0) {6} {\circle*{0.2}}
\multiput(-17.5,0) (1,0) {6} {\circle*{0.2}}
\setlength{\unitlength}{.8mm} 

\thicklines{
\put(19,0){\oval(28,19)[]}
\put(-19,0){\oval(28,19)[]}
}
\setlength{\unitlength}{.5mm} 
\put(2,-3){\makebox(0,0)[t]{{\protect\tiny 0}}}
\put(12,-3){\makebox(0,0)[t]{{\protect\tiny 1}}}

\put(29.5,-3){\makebox(0,0)[t]{{\protect\tiny  $p$--2}}}
\put(44,-4){\makebox(0,0)[t]{{\protect\tiny $p$--1}}}
\put(44,11){\makebox(0,0)[t]{{\protect\tiny $p$}}}
\put(-12,-3){\makebox(0,0)[t]{{\protect\tiny --1}}}

\put(-28,-3){\makebox(0,0)[t]{{\protect\tiny 2--${p}$}}}
\put(-45,-5){\makebox(0,0)[t]{{\protect\tiny 1--${p}$}}}
\put(-45,11){\makebox(0,0)[t]{{\protect\tiny --${p}$}}}
\put(35,-17){\makebox(0,0)[t]{{\protect\tiny $a(x),\bar{a}(x),G_{p-1}(x)$}}}
\put(-29,-17){\makebox(0,0)[t]{{\protect\tiny $a(x),\bar{a}(x),G_{p-1}(x)$}}}

\end{picture}}



\newsavebox{\kcgn}
\sbox{\kcgn}{\begin{picture}(140,25) (-70,-12.5)

\put(0,0){\circle*{3}}
\put(10,0){\circle{3}}
\put(20,0){\circle{3}}
\put(-10,0){\circle{3}}
\put(-20,0){\circle{3}}
\put(-30,0){\circle{3}}

\put(1.5,0){\line(1,0){7}}
\put(11.5,0){\line(1,0){7}}
\put(-8.5,0){\line(1,0){7}}
\put(-28.5,0){\line(1,0){7}}

\put(34,-2){$a$}


\multiput(22.5,0) (1,0) {6} {\circle*{0.2}}
\multiput(-17.5,0) (1,0) {6} {\circle*{0.2}}


\put(1.5,-3){\makebox(0,0)[t]{{\protect\scriptsize {\em k}}}}
\put(11.7,-3){\makebox(0,0)[t]{{\protect\scriptsize {\em k}+1}}}
\put(21.7,-3){\makebox(0,0)[t]{{\protect\scriptsize{\em k}+2}}}
\put(-8.5,-3){\makebox(0,0)[t]{{\protect\scriptsize {\em k}--1}}}
\put(-18.5,-3){\makebox(0,0)[t]{{\protect\scriptsize 2 }}}
\put(-28.5,-3){\makebox(0,0)[t]{{\protect\scriptsize  1 }}}

\end{picture}}



\newsavebox{\suz}
\sbox{\suz}{\begin{picture}(140,25) (-70,-12.5)

\put(0,0){\circle*{3}}
\put(10,0){\circle{3}}
\put(20,0){\circle{3}}
\put(-10,0){\circle{3}}
\put(-20,0){\circle{3}}
\put(-30,0){\circle{3}}

\put(1.5,0){\line(1,0){7}}
\put(11.5,0){\line(1,0){7}}
\put(-8.5,0){\line(1,0){7}}
\put(-28.5,0){\line(1,0){7}}


\multiput(22.5,0) (1,0) {6} {\circle*{0.2}}
\multiput(-17.5,0) (1,0) {6} {\circle*{0.2}}

\thicklines{
\put(17,0){\oval(36.5,15)[]}
}
\put(1.5,-3){\makebox(0,0)[t]{{\protect\scriptsize {\em k}}}}
\put(11.7,-3){\makebox(0,0)[t]{{\protect\scriptsize {\em k}+1}}}
\put(21.7,-3){\makebox(0,0)[t]{{\protect\scriptsize{\em k}+2}}}
\put(-8.5,-3){\makebox(0,0)[t]{{\protect\scriptsize {\em k}--1}}}
\put(-18.5,-3){\makebox(0,0)[t]{{\protect\scriptsize 2 }}}
\put(-28.5,-3){\makebox(0,0)[t]{{\protect\scriptsize  1 }}}
\put(19,-10){\makebox(0,0)[t]{{\protect\scriptsize  $a_0(x),\bar{a}_0(x),F(x)$ }}}

\end{picture}}


\newsavebox{\en}
\sbox{\en}{\begin{picture}(140,25) (-70,-12.5)

\put(0,0){\circle*{3}}
\put(10,0){\circle{3}}
\put(20,0){\circle{3}}
\put(30,0){\circle{3}}
\put(-10,0){\circle{3}}
\put(-20,0){\circle{3}}
\put(-30,0){\circle{3}}

\put(1.5,0){\line(1,0){7}}
\put(11.5,0){\line(1,0){7}}
\put(21.5,0){\line(1,0){7}}
\put(-8.5,0){\line(1,0){7}}
\put(-28.5,0){\line(1,0){7}}


\multiput(32.5,0) (1,0) {6} {\circle*{0.2}}
\multiput(-17.5,0) (1,0) {6} {\circle*{0.2}}

\thicklines{
\put(24,0){\oval(31,15)[]}
}
\put(1.5,-3){\makebox(0,0)[t]{{\protect\scriptsize {\em k}}}}
\put(13.7,-3){\makebox(0,0)[t]{{\protect\scriptsize {\em k}+1}}}
\put(23.7,-3){\makebox(0,0)[t]{{\protect\scriptsize{\em k}+2}}}
\put(33.7,-3){\makebox(0,0)[t]{{\protect\scriptsize{\em k}+3}}}
\put(-8.5,-3){\makebox(0,0)[t]{{\protect\scriptsize {\em k}--1}}}
\put(-18.5,-3){\makebox(0,0)[t]{{\protect\scriptsize 2 }}}
\put(-28.5,-3){\makebox(0,0)[t]{{\protect\scriptsize  1 }}}
\put(27,-10){\makebox(0,0)[t]{{\protect\scriptsize  $a(x),\bar{a}(x),F(x)$ }}}

\end{picture}}


\newsavebox{\suzen}
\sbox{\suzen}{\begin{picture}(140,25) (-70,-12.5)

\put(0,0){\circle*{3}}
\put(10,0){\circle{3}}
\put(20,0){\circle{3}}
\put(-10,0){\circle{3}}
\put(-20,0){\circle{3}}
\put(-30,0){\circle{3}}

\put(1.5,0){\line(1,0){7}}
\put(11.5,0){\line(1,0){7}}
\put(-8.5,0){\line(1,0){7}}
\put(-18.5,0){\line(1,0){7}}
\put(-28.5,0){\line(1,0){7}}

\put(38,-2){$a$}


\multiput(22.5,0) (1,0) {6} {\circle*{0.2}}
\multiput(-37.5,0) (1,0) {6} {\circle*{0.2}}

\thicklines{
\put(17,0){\oval(36.5,15)[]}
\put(-24,0){\oval(31,15)[]}
}
\put(1.5,-3){\makebox(0,0)[t]{{\protect\scriptsize {\em k}}}}
\put(11.7,-3){\makebox(0,0)[t]{{\protect\scriptsize {\em k}+1}}}
\put(21.7,-3){\makebox(0,0)[t]{{\protect\scriptsize{\em k}+2}}}
\put(-13.5,-3){\makebox(0,0)[t]{{\protect\scriptsize {\em k}--1}}}
\put(-21.5,-3){\makebox(0,0)[t]{{\protect\scriptsize {\em k}--2 }}}
\put(-30,-3){\makebox(0,0)[t]{{\protect\scriptsize {\em k}--3 }}}
\put(21,-10){\makebox(0,0)[t]{{\protect\scriptsize  $a_0(x),\bar{a}_0(x),F(x)$ }}}
\put(-21,-10){\makebox(0,0)[t]{{\protect\scriptsize  $a(x),\bar{a}(x),F(x)$ }}}

\end{picture}}


\newsavebox{\suzene}
\sbox{\suzene}{\begin{picture}(140,25) (-70,-12.5)

\put(0,0){\circle*{3}}
\put(10,0){\circle{3}}
\put(20,0){\circle{3}}
\put(30,0){\circle{3}}
\put(-10,0){\circle{3}}
\put(-20,0){\circle{3}}
\put(-30,0){\circle{3}}

\put(1.5,0){\line(1,0){7}}
\put(11.5,0){\line(1,0){7}}
\put(21.5,0){\line(1,0){7}}
\put(-8.5,0){\line(1,0){7}}
\put(-18.5,0){\line(1,0){7}}
\put(-28.5,0){\line(1,0){7}}

\put(42,-2){$b$}


\multiput(32.5,0) (1,0) {6} {\circle*{0.2}}
\multiput(-37.5,0) (1,0) {6} {\circle*{0.2}}

\thicklines{
\put(24,0){\oval(31,15)[]}
\put(-24,0){\oval(31,15)[]}
}
\put(1.5,-3){\makebox(0,0)[t]{{\protect\scriptsize {\em k}}}}
\put(14.7,-3){\makebox(0,0)[t]{{\protect\scriptsize {\em k}+1}}}
\put(23.4,-3){\makebox(0,0)[t]{{\protect\scriptsize{\em k}+2}}}
\put(-13.5,-3){\makebox(0,0)[t]{{\protect\scriptsize {\em k}--1}}}
\put(-20.5,-3){\makebox(0,0)[t]{{\protect\scriptsize {\em k}--2 }}}
\put(-30.5,-3){\makebox(0,0)[t]{{\protect\scriptsize {\em k}--3 }}}
\put(26,-10){\makebox(0,0)[t]{{\protect\scriptsize  $a(x),\bar{a}(x),F(x)$ }}}
\put(-22,-10){\makebox(0,0)[t]{{\protect\scriptsize  $a(x),\bar{a}(x),F(x)$ }}}

\end{picture}}



\newsavebox{\ofa}
\sbox{\ofa}{\begin{picture}(140,25) (-70,-12.5)

\put(0,0){\circle*{3}}
\put(10,0){\circle{3}}
\put(20,0){\circle{3}}
\put(-10,0){\circle{3}}
\put(-20,0){\circle{3}}

\put(1.5,0){\line(1,0){7}}
\put(11.5,0){\line(1,0){7}}
\put(-8.5,0){\line(1,0){7}}
\put(-18.5,0){\line(1,0){7}}

\put(33,-2){$b$}


\multiput(22.5,0) (1,0) {6} {\circle*{0.2}}
\multiput(-27.5,0) (1,0) {6} {\circle*{0.2}}



\put(1.5,-3){\makebox(0,0)[t]{{\protect\scriptsize {\em k}}}}
\put(11.7,-3){\makebox(0,0)[t]{{\protect\scriptsize {\em k}+1}}}
\put(21.7,-3){\makebox(0,0)[t]{{\protect\scriptsize{\em k}+2}}}
\put(-10,-3){\makebox(0,0)[t]{{\protect\scriptsize {\em k}--1}}}
\put(-20,-3){\makebox(0,0)[t]{{\protect\scriptsize {\em k}--2 }}}
\end{picture}}


\section{Introduction}

The SS-model is a two-parameter family of two-dimensional integrable quantum field theories (QFT), which possesses $U(1)\times U(1)$ symmetry and its action
can be written in terms of three boson fields with an exponential interaction \cite{SS}.
In this theory  the scattering matrix is the direct product of two S-matrices of the sine-Gordon (SG) model
 with different coupling constants. This family of models includes as particular cases some other interesting QFTs like the O(4)
 nonlinear sigma (NLS) model, the anisotropic principal chiral field and the $N=2$ supersymmetric SG model.
  
   The action of the model can be expressed by three boson fields  \cite{SS}:
  \begin{equation} 
  \mathcal{A}_{ss}=\int  d^2 x \ \left\{ 
  \frac12  \sum_{a=1}^3    ({\partial}_\mu \phi_a)^2+
  \frac{M_0}{\pi} \left[ \cos(\alpha  \phi_1+\tilde{\alpha}  \phi_2)e^{\beta \phi_3}+
  \cos(\alpha  \phi_1-\tilde{\alpha} \phi_2)e^{-\beta \phi_3} \right]
  \right\}, \label{1}
\end{equation}
where the parameters $\alpha, \tilde{\alpha}  $ and $\beta$ satisfy the relation :
\begin{equation}
\alpha^2+{\tilde{\alpha}  }^2-\beta^2=4 \pi .
\end{equation}
The S-matrix of the theory is :
\begin{equation}
S(\theta)=-S_{p}(\theta)\otimes S_{\tilde{p}}(\theta) \label{3},
\end{equation}
where $S_p(\theta)$ is the SG S-matrix with $\frac{\beta^2}{8 \pi}=\frac{p}{p+1}$.
 The relations between $p,\tilde{p}$ and the parameters of  action (\ref{1}) are :
 \begin{equation}
 p=\frac{{\alpha}^2}{2 \pi}, \qquad \tilde{p}=\frac{{\tilde{\alpha}  }^2}{2 \pi},\qquad  \frac{\beta^2}{2 \pi}=p+\tilde{p}-2.
 \end{equation}
  Action (\ref{1}) is unitary if $\beta^2>0$ so there are restrictions on the parameters $p,\tilde{p}$ :
 \begin{equation}
 p+\tilde{p} \geq 2.
 \end{equation}
 If $p \geq 1$ and $\tilde{p} \geq 1$  the spectrum of the model consist of four particles with equal masses.
If $p \rightarrow \infty$ and $\tilde{p} \rightarrow \infty$ then the S-matrix (\ref{3}) becomes $SU(2)\times SU(2)$ invariant
and it will coincide with the S-matrix of the $O(4)$ NLS model. Thus the SS-model can be regarded as a two-parameter
$U(1)\times U(1)$ invariant deformation of the $O(4)$ NLS model.

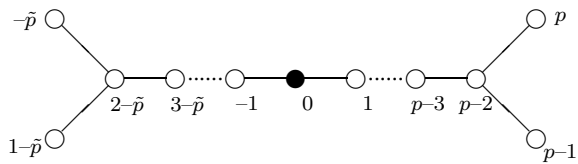
\begin{figure}[htbp]
\begin{center}
\begin{picture}(280,30)(-140,-15)
\put(-110,-10) {\usebox{\SSa}}

\put(-120,-19){\parbox{130mm}{\caption{ \label{1f}\protect {\small
Dynkin-diagram associated with the SS-model Y-system }}}}
\end{picture}
\end{center}
\end{figure}

 Once the spectrum and the associated S-matrix is known, the TBA method can be employed for calculating the ground state 
 energy of the model in finite volume. This was achieved  \cite{SS} for the  SS-model, and for integer values of $p$ and $\tilde{p}$
 the TBA equations can be encoded in an extended (affine) ${\mathcal{D}}_{p+\tilde{p}}$ Dynkin-diagram (see figure \ref{1f}.).
For general values of $p$ and $\tilde{p}$ the form of the integral equations are much more complicated, and the number of unknown 
functions depends on the continued fraction form of  $p$  and $\tilde{p}$  \cite{TS}. In the asymptotically free
case ($p,\tilde{p} \to \infty$) the number of unknown functions becomes infinite in the TBA equations. 
The solutions of the TBA equations also satisfy 
the well-known Y-system equations  \cite{ZamiY}.

 In  \cite{fi} it was noted that if a model has an S-matrix in the form of a direct product $S_G \otimes S_H$, and the TBA equations are already
 known for the models described by S-matrices $S_G$ and $S_H$, and the individual TBA equations are encoded on  Dynkin-like 
 diagrams of type $G$ and $H$ respectively, each having one massive node, then the TBA equations for the model with the
 direct product S-matrix can be obtained by gluing the individual TBA equations together at the massive node. This method can also be applied for the SS-model due to the tensor product form of the S-matrix.    
  
  Another approach for calculating finite size effects in a QFT is the use of some integrable lattice regularization
  of the QFT which can be solved by the Bethe Ansatz method.
   In models that can be solved by Bethe Ansatz, the well-known T- and Y-systems  \cite{Kuniba,KP} naturally appear and from them,
   using the analytical properties of the T-functions, one can easily derive the same TBA equations  which 
   can be obtained from the S-matrix.
    Using such an integrable regularization, an alternative method was worked out by Destri and de Vega  \cite{DdV0} for calculating
 finite size effects in the SG model. This is called the nonlinear integral equation (NLIE) technique. The advantage of this
 method is that one gets a single integral equation for any real value of the coupling constant, and the method can be extended
 to excited states, too  \cite{DdV2,DdV1,DdVres}. The problem is that this method can only be applied when the ground state of the model is
 formed by real Bethe Ansatz roots. In most  cases this is not true. For example the ground state of the spin-$S$  $XXX$-chain 
 is formed by $2S$-strings \cite{TB,pcm0,pcm1,KR}.
  
   Recently J. Suzuki \cite{suz} managed to derive some new types of nonlinear integral equations for describing the thermodynamics of
    the higher spin $XXX$-model at finite temperature, which can be regarded as the 
   generalization of the spin-$\frac12$ case which was treated successfully earlier  \cite{DdV0,KBP}. These new equations can be regarded
   as a particular mixture of TBA and NLIE, which gives back the $p\rightarrow \infty$ (isotropic) limit of the 
   Destri-de Vega equation  \cite{DdV0} in the spin-$\frac12 $ case.
    
	Last year motivated by the gluing idea  \cite{fi} and Suzuki's results  \cite{suz} Dunning proposed nonlinear integral equations  \cite{dun}
	similar to Suzuki's equations for describing the finite-volume ground state energy of some perturbed conformal field 
	theories whose S-matrices can be written in the form 
	of a direct product. These equations can also be regarded as a mixture of TBA and NLIE, but these equations have the 
	advantage that the number of unknown functions is less than in the TBA equations, hence these equations are much
	more convenient for numerical studies. Another advantage is that if the model has a single coupling constant then
	the equations  are valid for all real value of the coupling constant.

In this paper we derive a finite set of nonlinear integral equations for describing the finite size dependence of the
 ground state energy of the  asymptotically free $O(4)$ nonlinear sigma model. The equations are formulated in
 terms of two complex valued unknown functions and hence they are more convenient for numerical calculations
than the infinite set of TBA equations  \cite{SS,fon}.  
 By modifying the kernel functions of these equations we propose nonlinear integral equations for  the finite 
size effects in the SS-model for finite values of the couplings. The equations are also formulated in terms of two 
complex valued unknown functions and they are valid for arbitrary real values of the couplings. 
The form of these equations can be regarded as an appropriate gluing of two SG NLIEs together. 
	 
    The paper is organized as follows. In section 2, we recall the main results of the light-cone lattice approach to the $O(4)$ NLS model.
	In section 3, we introduce T- and Y-systems from the Bethe ansatz solution of the higher spin six-vertex model. 
	In section 4, using the light-cone lattice approach and  Suzuki's \cite{suz} method we derive nonlinear integral equations
	for describing the finite-volume ground state energy of the current-current perturbation of the $SU(2)_{2S}$ WZW model. 
    In section 5, using the light-cone lattice approach and motivated by Suzuki's method we derive new nonlinear 
	integral equations for the same model.
    In section 6,  in the $S\rightarrow \infty$ limit we glue together the two nonlinear integral equations to a much 
	simpler equation. This equation describes the finite-volume ground state energy of the $O(4)$ NLS model.
	In section 7, we propose nonlinear integral equations for describing the finite-volume ground state energy of 
	the SS-model by modifying the kernel functions of the equations obtained for the $O(4)$ NLS model.
	In section 8, we perform some analytical and numerical tests on the proposed equations.
	In section 9, some special cases are considered.
	The summary and conclusions of this paper are given in section 10.

\section{The light-cone lattice approach to the $O(4)$ NLS model}

Our starting point is the Bethe Ansatz solution of the integrable lattice
regularization of the $O(4)$ NLS model \cite{LC,LCS}. In this section we briefly
summarize the results of this approach \cite{LCS}. The fields of the regularized theory
are defined at sites (\lq\lq events") of a light-cone lattice and the
dynamics of the system is defined by 
translations in the left and right light-cone directions. These are given by
transfer matrices of the isotropic higher spin six-vertex model  with
alternating inhomogeneities. This approach is particularly useful for
calculating the finite size dependence of physical quantities.
We take $N$ points ($N$ is even) in the spatial direction and use periodic
boundary conditions. The lattice spacing is related to $l$, the
(dimensionful) size of the system :
\begin{equation}
a=\frac{2l}{N}.
\end{equation}

The physical states of the system are characterized by the set of
Bethe roots \break $\{x_j,\ j=1,\dots,M \}$, which satisfy the 
Bethe Ansatz equations (BAE) 
\begin{equation}
\left\{   \frac{(x_j+x_0+ik)(x_j-x_0+ik)}{(x_j+x_0-ik)(x_j-x_0-ik) } 
\right\}^{\frac{N}{2}}=- \frac{Q(x_j+2i)}{Q(x_j-2i)}
\qquad \qquad j=1,..,M
\end{equation}
where 
\begin{equation}
k=2S, \qquad \qquad Q(x)=\prod_{j=1}^{M}(x-x_j),
\end{equation} 
and $x_0$ is the inhomogeneity parameter. 

  The energy ($E$) and momentum ($P$) of the physical states can be
obtained from the eigenvalues of the light-cone transfer matrices:
\begin{equation}
e^{i \frac{a}{2}(H+P)}=(-1)^M \frac{Q(x_0-ik)}{Q(x_0+ik)}, 
\qquad \qquad 
e^{i \frac{a}{2}(H-P)}=(-1)^M \frac{Q(-x_0+ik)}{Q(-x_0-ik)}. \label{9}
\end{equation}
Besides the usual procedure, taking the thermodynamic limit
($N \to \infty$) first, followed by the continuum limit ($a \to 0$)
one can also study continuum limit in finite volume by taking
$N \to \infty$ and tuning the inhomogeneity parameter $x_0$
simultaneously as
\begin{equation}
x_0=\frac{2}{\pi}\log \frac{4}{{m}a}=\frac{2}{\pi}\log \frac{2N}{ml},
\label{10}
\end{equation}
where the mass parameter ${m}$ is the
infinite volume mass gap of the theory. If we take this continuum limit at a finite, fixed spin value $S$ 
we get the energy and momentum eigenvalues of  the current-current perturbation of the $SU(2)_{2S}$
WZW model. After this continuum limit we shall take the $S \to \infty$ limit in order to obtain the energy and momentum
eigenvalues of the $O(4)$ NLS model in finite volume with periodic boundary conditions  \cite{LCS}.
 
 It is important to realize that
this procedure {\em does not} give all the eigenvalues of the Hamilton and the momentum operators of the $O(4)$ NLS model!
Representing the $O(4)$ symmetry of the model as 
$O(4) \simeq SU(2)_L \times SU(2)_R$, one can show that only those 
states of the Hilbert space of the theory can be described by this method which are $SU(2)_L$ (or equivalently $SU(2)_R$)
 singlets  \cite{LCS,pcm0}. Since the ground
state of the $O(4)$ NLS model lies in this sector of the Hilbert space,  this description is appropriate to examine
its finite size dependence. 
We note that the same results could be obtained for the $O(4)$ NLS model by using the approach
of  refs. \cite{pcm0,pcm1}.


\section{ Fusion hierarchy and T-systems}

In this section we introduce  T-systems from the fusion hierarchy of the isotropic spin-$S$ six-vertex model.
Let $V_i \simeq \mathbb{C}^{l_i+1}$  be the irreducible $SU(2)$ representation with spin $l_i/2$, and let 
$R^{(l_i,l_j)}_{ij}(x)$ be a linear operator acting on $V_i \otimes V_j$. These operators depend on a complex parameter
 $x$ and satisfy the Yang-Baxter equation
\begin{equation}
R^{(l_1,l_2)}_{12}(x) R^{(l_1,l_3)}_{13}(x+y) R^{(l_2,l_3)}_{23}(y)=
R^{(l_2,l_3)}_{23}(y) R^{(l_1,l_3)}_{13}(x+y) R^{(l_1,l_2)}_{12}(x). \label{11}
\end{equation}
The solutions of this equation can be obtained by fusion \cite{fusi} from the simplest R-matrix which corresponds
to the spin-$\frac12$ representation both in $V_1$ and $V_2$. This R-matrix is of the form
\begin{equation}
R^{(1,1)}(x)=x+i+i \sum_{a=1}^{3} \sigma^a \otimes \sigma^a,
\end{equation}
where  $\sigma^a$ are the Pauli matrices.  The concrete form of the $R^{(l_i,l_j)}(x)$ matrices 
can be taken from ref. \cite{KR}. 
From these R-matrices one can define the family of monodromy matrices with
alternating inhomogeneities ${x_i=(-1)^{i+1} x_0}$:   
\begin{equation}
T^{(p)}(x,\{x_i\})= R^{(p,k)}_{a1}(x-x_1-i(k+p-1)) \dots R^{(p,k)}_{aN}(x-x_N-i(k+p-1)), 
\end{equation}
where $k=2S$.
These matrices act on the tensor product $V_a \otimes V_H$ where $V_H=V_1 \otimes ... \otimes V_N$ is the
quantum space of the system and $V_a$ is the auxiliary space with  $ V_i \simeq \mathbb{C}^k, V_a \simeq \mathbb{C}^{\, p}$.
In the rest of the paper we consider the case when $N$ is even, because this is necessary for the light-cone lattice approach. 
Define the transfer matrix by taking the trace of the monodromy matrix over the auxiliary space; 
\begin{equation}
\tau_p (x,\{x_i\})=\mbox{Tr}_a T^{(p)}(x,\{x_i\}). \label{14}
\end{equation}
Due to the Yang-Baxter relation (\ref{11})  the transfer matrices (\ref{14}) form a commutative family of operators acting on
$V_H$ \cite{com,aba}
\begin{equation}
\left[ \tau_p (x,\{x_i\}),\tau_n (y,\{x_i\}) \right]=0.
\end{equation}
Due to the integrability and commutativity all these transfer matrices can be simultaneously diagonalized by the
algebraic Bethe Ansatz \cite{aba}. The eigenvalues of the transfer matrices can be characterized by the solutions of the
Bethe Ansatz equations \cite{KR}
\begin{equation}
\left\{   \frac{(x_j+x_0+ik)(x_j-x_0+ik)}{(x_j+x_0-ik)(x_j-x_0-ik) } 
\right\}^{\frac{N}{2}}=- \frac{Q(x_j+2i)}{Q(x_j-2i)}
\qquad \qquad j=1,..,M. \label{16}
\end{equation}
The eigenvalues of the transfer matrices (\ref{14}) are of the form \cite{KR}
\begin{equation}
T_p(x)=\sum_{l=1}^{p+1} \  \psi^{(p)}_{l} (x)  \  \frac{Q(x+i(p+1)) \ Q(x-i(p+1)) }{Q(x+i(2l-p-1)) \ Q(x+i(2l-p-3)) },
\label{17}
\end{equation}
where 
 \begin{equation}
\psi^{(p)}_{l} (x)= \prod_{m=l-1}^{p-1} \phi( x+2im-i(k+p-1)) \prod_{n=0}^{l-2} \phi(x+2in+i(k+1-p)),
\end{equation}
\begin{equation}
\phi(x)=(x-x_0)^{\frac{N}{2}}(x+x_0)^{\frac{N}{2}}.
\end{equation}
The eigenvalues (\ref{17}) of the transfer matrices (\ref{14}) satisfy  the so-called T-system equations:
\begin{equation}
T_p(x+i) T_p(x-i)= f_p(x)+T_{p-1}(x) T_{p+1}(x), \label{20}
\end{equation}
where 
\begin{equation}
f_p(x)=\prod_{j=1}^{p} \phi(x+i(2j-p+k)) \ \phi(x+i(2j-p-k-2)).  
\end{equation}
From a T-system (\ref{20}) one can define a Y-system as follows
\begin{equation}
y_j(x)=\frac{T_{j-1}(x) T_{j+1}(x) }{f_j(x)}, \label{22}
\end{equation}
\begin{equation}
Y_j(x)=1+y_j(x)=\frac{ T_j (x+i) T_j (x-i) }{ f_j(x) }. \label{23}
\end{equation}
These functions satisfy the Y-system equations \cite{ZamiY,Kuniba,KP}
\begin{equation}
y_j(x+i) y_j(x+i)=Y_{j-1}(x) Y_{j+1}(x).
\end{equation}
One can redefine the T-system elements with
\begin{equation}
T_j(x) \rightarrow \tilde{T}_j(x)=\sigma_j(x) T_j(x),
\end{equation}
where $\sigma_j(x)$ satisfies the relation
\begin{equation}
\sigma_j(x+i) \sigma_j(x-i)=\sigma_{j-1}(x) \sigma_{j+1}(x),
 \end{equation}
then the new $\tilde{T}_j(x)$ functions satisfy the same T-system relations as (\ref{20}) but with different $f_j(x)$
functions: 
\begin{equation}
f_j(x) \rightarrow \tilde{f}_j(x)=\sigma_j(x+i) \sigma_j(x-i) f_j(x).
\end{equation}
This is called gauge transformation. Under such a transformation the Y-system (\ref{22}-\ref{23}) is invariant.
As we will see later the energy of  the model can be expressed by an element of the gauge invariant Y-system.
On the other hand one has the freedom of choosing that gauge for the T-system (\ref{20}) which is convenient for 
the particular calculations.


\section{Nonlinear integral equations I.}

In this section we recall the derivation of Suzuki's equations  \cite{suz}
 to describe the ground state energy of our model in finite
volume. Consider the following gauge transformation of the T-system (\ref{20})
\begin{equation}
\tilde{T}_p(x+i)\tilde{T}_p(x-i)=\tilde{f}_p (x)+\tilde{T}_{p-1}(x)\tilde{T}_{p+1}(x), \label{28}
\end{equation}
where
\begin{equation}
\tilde{f}_p(x)=\prod_{j=1}^{p} \phi(x+i(p-k-2j)) \ \phi(x-i(p-k-2j)),
\end{equation}  
and
\begin{equation}
\tilde{T}_{-1}(x)=0, \qquad \qquad \tilde{T}_0(x)=1.
\end{equation}
The solutions of these equations are  of the form \cite{suz} 
\begin{equation}
\tilde{T}_p(x)=\sum_{l=1}^{p+1} \tilde{\lambda}_{l}^{(p)}(x),
\end{equation}
where
\begin{equation}
\tilde{\lambda}_{l}^{(p)}(x)=\tilde{\psi}_{l}^{(p)}(x)\ \frac{Q(x+i(p+1)) \ Q(x-i(p+1)) }{Q(x+i(2l-p-1)) \ Q(x+i(2l-p-3)) },
\end{equation}
\begin{equation}
\tilde{\psi}_{l}^{(p)}(x) = \prod_{j=1}^{p-l+1} \phi(x+i(p-k-2j+1)) \prod_{j=1}^{l-1} \phi(x-i(p-k-2j+1)). \label{33}
\end{equation}
We define the following auxiliary functions \cite{suz}
\begin{equation}
y_j(x)=\frac{\tilde{T}_{j-1}(x) \tilde{T}_{j+1}(x) }{\tilde{f}_j(x)}, \ \qquad \qquad j=1,\dots,k \label{34}
\end{equation}
\begin{equation}
Y_j(x)=1+y_j(x)=\frac{ \tilde{T}_j (x+i) \tilde{T}_j (x-i) }{\tilde{ f}_j(x) }, \ \qquad \qquad j=1,\dots,k \label{35}
\end{equation}
\begin{equation}
b(x)=\frac{ \tilde{\lambda}_{1}^{(k)}(x+i)+\dots+\tilde{\lambda}_{k}^{(k)}(x+i) }{\tilde{\lambda}_{k+1}^{(k)}(x+i) },
\qquad \qquad \mathcal{B}(x)=1+b(x),
\end{equation}
\begin{equation}
\bar{b}(x)=\frac{ \tilde{\lambda}_{2}^{(k)}(x-i)+\dots+\tilde{\lambda}_{k+1}^{(k)}(x-i) }{\tilde{\lambda}_{1}^{(k)}(x-i) },
\qquad \qquad \bar{\mathcal{B}}(x)=1+\bar{b}(x). \label{37}
\end{equation}
From (\ref{28}-\ref{33}) it follows that the auxiliary functions (\ref{34}-\ref{37}) satisfy the following functional relations:
\begin{equation}
\tilde{T}_k(x+i)=\prod_{j=1}^{k} \phi(x+2ij) \ \frac{Q(x-ik) }{ Q(x+ik) } \ \mathcal{B}(x), \label{38}
\end{equation}
\begin{equation}
\tilde{T}_k(x-i)=\prod_{j=1}^{k} \phi(x-2ij) \ \frac{Q(x+ik) }{ Q(x-ik) } \ \bar{\mathcal{B}}(x), \label{39}
\end{equation}
\begin{equation}
b(x)=\frac{ \phi(x) }{ \prod_{j=1}^{k} \phi(x+2ij) } \  \frac{ Q(x+ik+2i) }{ Q(x-ik) } \ \tilde{T}_{k-1}(x), \label{40}
\end{equation}
\begin{equation}
\bar{b}(x)=\frac{ \phi(x) }{ \prod_{j=1}^{k} \phi(x-2ij) } \  \frac{ Q(x-ik-2i) }{ Q(x+ik) } \ \tilde{T}_{k-1}(x), \label{41}
\end{equation}
\begin{equation}
\mathcal{B}(x) \bar{\mathcal{B} }(x)=Y_k(x), \label{42}
\end{equation}
\begin{equation}
y_j(x+i) y_j(x-i)=Y_{j-1}(x) Y_{j+1}(x) \qquad \qquad j=1,\dots,k-2, \label{43}
\end{equation}
\begin{equation}
y_{k-1}(x+i) y_{k-1}(x+i) = Y_{k-2}(x) \mathcal{B}(x) \bar{\mathcal{B} }(x),
\end{equation}
\begin{equation}
\tilde{T}_{k-1}(x+i) \tilde{T}_{k-1}(x-i)=\tilde{f}_{k-1}(x) Y_{k-1}(x). \label{45}
\end{equation}
We introduce two other auxiliary functions
\begin{equation}
\Psi_1(x)=Q(x-ik), \qquad \qquad \Psi_2(x)=Q(x+ik). \label{46}
\end{equation}
In order to be able to derive integral equations from these functional relations one needs to know the positions 
of the zeroes and the poles of the auxiliary functions (\ref{34}-\ref{37}). Due to the relations (\ref{38}-\ref{45}) we only 
need to know the zeroes of $Q(x)$ and $\tilde{T}_j(x)$. For the ground state of our model the zeroes of  $Q(x)$ form
 $\frac{N}{2}$ pieces  of $k$-strings, but for  finite $N$ values there are deviations from the string hypothesis \cite{str}. 
From \cite{str} one can see that those Bethe roots have the smallest deviations from the imaginary value of the string
hypothesis that have imaginary parts $\pm (k-1)$ according to the string hypothesis. These deviations are always less
than $1/2$ and this is important 
 because the analytic properties of $\Psi_1(x)$ ($\Psi_2(x)$) are influenced mainly by these
roots on the upper (lower) half plane of the complex plane near the real axis. 

\begin{figure}[htbp]
\begin{center}
\begin{picture}(280,30)(-140,-15)
\put(-120,-10) {\usebox{\suz}}

\put(-130,-15){\parbox{130mm}{\caption{ \label{2f}\protect {\small
Graphical representation of nonlinear integral equations I. }}}}
\end{picture}
\end{center}
\end{figure}
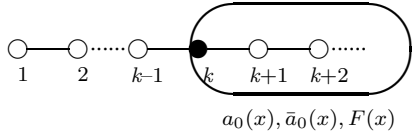

  The transfer matrices $\tilde{T}_j(x) \quad j=1,\dots,k$ 
(\ref{28}) have no zeroes in the ground state in the \lq\lq main" strip $0\leq|\mbox{Im}x | \leq 1$.  We can list the strips
where the auxiliary functions are analytic and non zero (ANZ).
\begin{eqnarray}
\Psi_1(x) \qquad  &\mbox{ANZ}& \qquad \mbox{Im}\ x \leq 1/2,   \nonumber \\
\Psi_2(x) \qquad &\mbox{ANZ}& \qquad \mbox{Im}\ x \geq -1/2,  \nonumber  \\
b(x),\mathcal{B}(x) \qquad &\mbox{ANZ}& \qquad 0 < |\mbox{Im}\ x| \leq 1/2,   \nonumber  \\ 
\bar{b}(x),\bar{\mathcal{B}}(x) \qquad &\mbox{ANZ}& \qquad 0 < |\mbox{Im}\ x| \leq 1/2, \label{47}\\
y_j(x)  \qquad &\mbox{ANZ}& \qquad 0 \leq |\mbox{Im}\ x| \leq 1, \qquad j=1,\dots,k-1,   \nonumber  \\
Y_j(x) \qquad &\mbox{ANZ}& \qquad 0 \leq |\mbox{Im}\ x| \leq \epsilon, \qquad j=1,\dots,k-1 \quad \epsilon>0,   \nonumber  \\
\tilde{T}_j(x)  \qquad &\mbox{ANZ}& \qquad 0 \leq |\mbox{Im}\ x| \leq 1, \qquad j=1,\dots,k. \nonumber 
\end{eqnarray}
We introduce new variables by shifting the arguments of $b(x),\mathcal{B}(x)$ and $\bar{b}(x),\bar{\mathcal{B}}(x)$ by $\pm i \gamma$ \cite{suz} 
\begin{equation}
a_0(x)=b(x-i\gamma), \qquad \qquad U_0(x)=\mathcal{B}(x-i\gamma)=1+a_0(x),
\end{equation}
 \begin{equation}
\bar{a}_0(x)=\bar{b}(x+i\gamma), \qquad \qquad 
\bar{U}_0(x)=\bar{\mathcal{B}}(x+i\gamma)=1+\bar{a}_0(x),
\end{equation}
where $0<\gamma<1/2$ is an arbitrary real parameter. This shift is necessary because the original functions
$b(x),\bar{b}(x)$ have zeroes on the real axis.
Due to the ANZ property of $\tilde{T}_k(x)$ (\ref{47}) and the fact that $\lim_{|x| \to \infty} \frac{d}{dx} \log  \tilde{T}_k(x) = 0$ the following relation holds \cite{suz}
\begin{equation}
0=\int\limits_{-\infty}^{\infty}dx \ \frac{d}{dx} \log  \tilde{T}_k(x-i)\ e^{iq(x-i)}-\int\limits_{-\infty}^{\infty} dx \
\frac{d}{dx} \log  \tilde{T}_k(x+i)\ e^{iq(x+i)}. \label{50}
\end{equation}
From (\ref{38}) and (\ref{39}) one can express $\tilde{T}_k(x \pm i)$ with the auxiliary functions and by substituting these 
expressions  into (\ref{50}) one gets  in  Fourier space (see conventions for Fourier transformation in appendix A.)
\begin{eqnarray}
\tilde{dl} \Psi_1 (q>0) &=& \pi i N e^{-kq} \ \frac{ \sinh (kq) \ \cos (x_0 q) }{ \cosh (q) \ \sinh (q) } \nonumber \\
                                      &+& \frac{ e^{(1-\gamma)q } }{2 \cosh (q) } \ \tilde{dl} \bar{U}_0 (q) - 
                                                  \frac{ e^{-(1-\gamma)q } }{2 \cosh (q) } \ \tilde{dl} {U}_0 (q),  \\
\tilde{dl} \Psi_1 (q<0) &=& 0,  \\
\tilde{dl} \Psi_2 (q>0) &=& 0 ,   \\
\tilde{dl} \Psi_2 (q<0) &=& -\pi i N e^{kq} \ \frac{ \sinh (kq) \ \cos (x_0 q) }{ \cosh (q) \ \sinh (q) } \nonumber \\
                                      &-& \frac{ e^{(1-\gamma)q } }{2 \cosh (q) } \ \tilde{dl} \bar{U}_0 (q) + 
                                                  \frac{ e^{-(1-\gamma)q } }{2 \cosh (q) } \ \tilde{dl} {U}_0 (q), 
\end{eqnarray}
where we introduced the notation
\begin{equation}
\tilde{dl} F(q)=\int\limits_{-\infty}^{\infty} dx \ e^{iqx} \ \frac{d}{dx} \log F(x).
\end{equation}
One can derive similar relations for $\tilde{dl} y_j(q)$'s and  $\tilde{dl} Y_j(q)$'s from (\ref{43}), and $\tilde{dl} \tilde{T}_{k-1}(q)$
and $\tilde{dl} Y_{k-1}(q)$ from (\ref{45}). Substituting these relations into the definitions of $a_0(x)$ and $\bar{a}_0(x)$, one obtains $k+1$ algebraic relations in Fourier space. After taking the inverse Fourier
transformation of these relations and integrating over $x$, we get Suzuki's hybrid nonlinear integral equations
 \cite{suz}
\begin{eqnarray}
\log y_1(x) &=& (K*\log Y_2)(x), \nonumber \\  
\log y_j (x) &=& (K* \log Y_{j-1})+(K*\log Y_{j+1})(x), \qquad  \quad j=2,\dots,k-2,  \nonumber  \\   
\log y_{k-1}(x) &=&  (K*\log Y_{k-2})(x)+ (K^{+\gamma}*\log U_0)(x)+
                                         (K^{-\gamma}*\log \bar{U}_0)(x),    \nonumber  \\
\log a_0(x) &=& \mathcal{D}_N (x-i \gamma)+ (F*\log  U_0)(x)- (F^{+2(1-\gamma)}*\log \bar{U}_0 )(x)  \label{58} \\
                            &+& (K^{-\gamma}*\log Y_{k-1})(x),    \nonumber  \\
\log \bar{a}_0(x) &=& \mathcal{D}_N (x+i \gamma)+ (F*\log  \bar{U}_0)(x)- (F^{-2(1-\gamma)}*\log  U_0 )(x)    \nonumber  \\
                             &+& (K^{+\gamma}*\log Y_{k-1})(x),   \nonumber
\end{eqnarray}
where $(K*f)(x)=\int dy \ K(x-y)\ f(y) $ is the convolution and the \lq\lq source" function $\mathcal{D}_N (x)$ on the lattice 
reads as:
\begin{equation}
\mathcal{D}_N (x)= i N \arctan \left[    \frac{ \sinh \left(  \frac{\pi (x+i)}{2}   \right)  }{ \cosh \left(   \frac{\pi x_0}{2}   \right)               }        \right] ,\qquad \qquad x_0=\frac{2}{\pi} \log \left(    \frac{2N}{ml}  \right).
\end{equation}
The kernel functions of (\ref{58}) are of the form 
\begin{equation}
K(x)=\frac{1}{4 \ \cosh (\pi x /2) }, \label{60}
\end{equation}
\begin{equation}
F(x)=\int\limits_{-\infty}^{\infty} \frac{dq}{2 \pi} \  \frac{e^{-|q|-iqx}}{2 \cosh(q) },
\end{equation}
and we have used the notation
\begin{equation}
f^{\pm \eta}(x)=f(x \pm i \eta).
\end{equation}
We get the continuum limit of eqs. (\ref{58}) by taking the $N \to \infty$ limit. The kernel functions
do not change because they are independent of $N$, but the \lq\lq source" function changes  in the continuum as
\begin{equation}
\mathcal{D}(x)=\lim_{N \to \infty} \mathcal{D}_N (x)=-ml \cosh \left( \frac{\pi x}{2} \right). \label{63}
\end{equation}
In the TBA language the complex auxiliary functions $a_0(x)$ and $\bar{a}_0(x)$ resum the contributions of 
those Y-system elements whose index is larger than $k-1$. Equations (\ref{58}) are graphically represented in figure \ref{2f}.
 The big \lq\lq bubble" denotes the complex auxiliary functions which resum the contributions of those
TBA nodes, which are inside it. In our notation the names of the complex unknown
functions and the kernel function are indicated.

The energy and momentum of the model (\ref{9}) can be easily expressed by our auxiliary functions due to the fact
that  apart from some trivial normalization factors: 
\begin{equation}
e^{i \frac{a}{2}(H+P)} \sim \tilde{T}_k(x_0+i),  \qquad \mbox{and} \qquad e^{i \frac{a}{2}(H-P)} \sim \tilde{T}_k(-x_0-i).
\nonumber
\end{equation}
After some straightforward calculations \cite{Kuniba,SGg} one gets in the continuum limit for the ground state energy of the model 
\begin{eqnarray}
E_0(l) &=& E_{bulk}-\frac{m}{4} \int\limits_{-\infty}^{\infty} dx  \ \cosh \left( \frac{\pi (x-i \gamma)}{2}\right) \log U_0(x) \label{65} \\
&-& \frac{m}{4} \int\limits_{-\infty}^{\infty} dx \
\cosh \left( \frac{\pi (x+i \gamma)}{2}\right) \log \bar{U}_0(x) , \nonumber
 \end{eqnarray}
where $E_{bulk}$ is an overall divergent factor in the continuum and it has the form on the lattice:
\begin{equation}
E_{bulk}=\frac{N^2}{2l} \ \sum_{j=0}^{k/2-1} \ \frac{1}{i} \ \log\left[  \frac{i(1+2j)-x_0}{i(1+2j)+x_0}  \right]
\qquad  \qquad \mbox{when}\  k \ \mbox{is even,}
\end{equation}
\begin{equation}
E_{bulk}=\frac{N^2}{2l} \left(  \chi(2x_0) +  
\sum_{j=0}^{(k-3)/2} \ \frac{1}{i} \ \log\left[  \frac{2i(1+j)-x_0}{2i(1+j)+x_0}  \right]  \right)
\qquad  \quad \mbox{when}\ k \ \mbox{is odd,}
\end{equation}
where
\begin{equation}
\chi(x)=  \int\limits_{-\infty}^{\infty} \frac{dq}{2 \pi} \  \frac{\sin(qx)}{q}
\frac{e^{-|q|}}{2 \cosh(q) }.
\end{equation}
From these formulas one can easily see that $E_{bulk}$ is independent of the Bethe roots, thus independent of the dynamics
of the model and so $E_{bulk}$ appears in the energy expression as a bulk energy term same for all states of the model. 
One can easily see that $E_{bulk}$ is divergent in the continuum, but after subtracting this divergent term from the energy  
the finite integral terms in (\ref{65}) describe the finite size dependence of the ground state energy of the model.


\section{Nonlinear integral equations II.}

\begin{figure}[htbp]
\begin{center}
\begin{picture}(280,30)(-140,-15)
\put(-120,-10) {\usebox{\en}}

\put(-130,-15){\parbox{130mm}{\caption{ \label{3f}\protect {\small
Graphical representation of nonlinear integral equations II.  }}}}
\end{picture}
\end{center}
\end{figure}
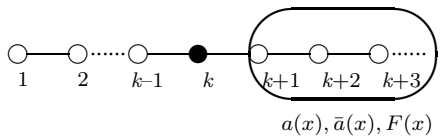

In this section we derive a new set of nonlinear integral equations to describe the ground state energy of our model
 in finite volume. Although these equations are  similar to Suzuki's equations they differ in that the
\lq\lq source" term is coupled to the TBA part of the system
and in the TBA language the complex auxiliary functions resum the contributions of those Y-system elements 
whose index is larger than $k$. (See figure \ref{3f}.)
Consider another  gauge transformation  of  T-system (\ref{20})
\begin{equation}
\hat{T}_p(x+i)\hat{T}_p(x-i)=\hat{f}_p (x)+\hat{T}_{p-1}(x)\hat{T}_{p+1}(x), \label{69}
\end{equation}
where
\begin{equation}
\hat{f}_p(x)=\hat{T}_0(x+i(p+1)) \ \hat{T}_0(x-i(p+1)),
\end{equation}  
and
\begin{equation}
\hat{T}_{-1}(x)=0, \qquad \qquad \hat{T}_0(x)=\prod_{j=0}^{k-1} \phi(x+i(k-1-2j)).
\end{equation}
The solutions of these equations are of the form
\begin{equation}
\hat{T}_p(x)=\sum_{l=1}^{p+1} \hat{\lambda}_{l}^{(p)}(x),
\end{equation}
where
\begin{equation}
\hat{\lambda}_{l}^{(p)}(x)=\hat{T}_0(x+i(2l-p-2)) \ \frac{Q(x+i(p+1)) \ Q(x-i(p+1)) }{Q(x+i(2l-p-1)) \ Q(x+i(2l-p-3)) }. \label{73}
\end{equation}
We define the following auxiliary functions: 
\begin{equation}
y_j(x)=\frac{\hat{T}_{j-1}(x) \hat{T}_{j+1}(x) }{\hat{f}_j(x)}, \ \qquad \qquad j=1,\dots,k+1, \label{74}
\end{equation}
\begin{equation}
Y_j(x)=1+y_j(x)=\frac{ \hat{T}_j (x+i) \hat{T}_j (x-i) }{\hat{ f}_j(x) }, \ \qquad \qquad j=1,\dots,k+1, \label{75}
\end{equation}
\begin{equation}
h(x)=\frac{ \hat{\lambda}_{1}^{(k+1)}(x+i)+\dots+\hat{\lambda}_{k+1}^{(k+1)}(x+i) }{\hat{\lambda}_{k+2}^{(k+1)}(x+i) },
\qquad \qquad \mathcal{H}(x)=1+h(x), \label{76}
\end{equation}
\begin{equation}
\bar{h}(x)=\frac{ \hat{\lambda}_{2}^{(k+1)}(x-i)+\dots+\hat{\lambda}_{k+2}^{(k+1)}(x-i) }{\hat{\lambda}_{1}^{(k+1)}(x-i) },
\qquad \qquad \bar{\mathcal{H}}(x)=1+\bar{h}(x). \label{77}
\end{equation}
From (\ref{69}-\ref{73}) it follows that the auxiliary functions (\ref{74}-\ref{77}) satisfy the following functional relations:
\begin{equation}
\hat{T}_{k+1}(x+i)=\hat{T}_0(x+i(k+2)) \ \frac{Q(x-ik-i) }{ Q(x+ik+i) } \ \mathcal{H}(x), \label{78}
\end{equation}
\begin{equation}
\hat{T}_{k+1}(x-i)=\hat{T}_0(x-i(k+2)) \ \frac{Q(x+ik+i) }{ Q(x-ik-i) } \ \bar{\mathcal{H}}(x),\label{79}
\end{equation}
\begin{equation}
h(x)=\frac{1}{\hat{T}_0(x+i(k+2))} \  \frac{ Q(x+ik+3i) }{ Q(x-ik-i) } \ \hat{T}_{k}(x),
\end{equation}
\begin{equation}
\bar{h}(x)= \frac{1}{\hat{T}_0(x-i(k+2))}\  \frac{ Q(x-ik-3i) }{ Q(x+ik+i) } \ \hat{T}_{k}(x),
\end{equation}
\begin{equation}
\mathcal{H}(x) \bar{\mathcal{H} }(x)=Y_{k+1}(x), \label{82}
\end{equation}
\begin{equation}
y_j(x+i) y_j(x-i)=Y_{j-1}(x) Y_{j+1}(x) \qquad \qquad j=1,\dots,k-1,
\end{equation}
\begin{equation}
y_{k}(x+i) y_{k}(x+i) = Y_{k-1}(x) \mathcal{H}(x) \bar{\mathcal{H} }(x),
\end{equation}
\begin{equation}
\hat{T}_{k}(x+i) \hat{T}_{k}(x-i)=\hat{T}_0(x+i(k+1)) \ \hat{T}_0(x-i(k+1))  Y_{k}(x). \label{85}
\end{equation}
In order to be able to derive integral equations from these functional relations one needs to know the positions of the
zeroes and the poles of the auxiliary functions (\ref{74}-\ref{77}). Due to the relations (\ref{78}-\ref{85}) we only need
  the zeroes of 
$Q(x)$ and $\hat{T}_j(x)$. The zeroes of $Q(x)$ are the same as in the previous section, but the 
$\hat{T}_j(x)$s have changed due to the gauge transformation. This change is only an overall  factor which 
does not depend on the Bethe roots. So in this gauge all $\hat{T}_j(x)$ can have  $\frac{N}{2}$-fold degenerate 
zeroes in  the \lq\lq main" strip at places $\pm x_0, \pm x_0 \pm i$, but these zeroes cancel from the Y-system 
elements (\ref{74}-\ref{75}) except the $y_k(x)$ case which has $\frac{N}{2}$-fold degenerate zeroes in the \lq\lq main" strip at $\pm x_0$.
 These zeroes will give the standard TBA source term in our final equations \cite{Kuniba,SGg}.
 Now we can list the strips,
where the auxiliary functions are analytic and non zero (ANZ):
\begin{eqnarray}
\Psi_1(x) \qquad  &\mbox{ANZ}& \qquad \mbox{Im}\ x \leq 1/2,   \nonumber  \\ 
\Psi_2(x) \qquad &\mbox{ANZ}& \qquad \mbox{Im}\ x \geq -1/2,   \nonumber  \\
h(x),\mathcal{H}(x) \qquad &\mbox{ANZ}& \qquad -3/2 < \mbox{Im}\ x \leq 0,   \nonumber  \\
\bar{h}(x),\bar{\mathcal{H}}(x) \qquad &\mbox{ANZ}& \qquad 0 \leq \mbox{Im}\ x < 3/2, \label{86}\\
y_j(x)  \qquad &\mbox{ANZ}& \qquad 0 \leq |\mbox{Im}\ x| \leq 1, \qquad j=1,\dots,k-1,    \nonumber  \\ 
Y_j(x) \qquad &\mbox{ANZ}& \qquad 0 \leq |\mbox{Im}\ x| \leq \epsilon, \qquad j=j,\dots,k \quad \epsilon>0,   
\nonumber  \\
\hat{T}_{k+1}(x)  \qquad &\mbox{ANZ}& \qquad 0 \leq |\mbox{Im}\ x| \leq 1 . \nonumber 
\end{eqnarray}
We introduce new variables by shifting the arguments of $h(x),\mathcal{H}(x)$ and $\bar{h}(x),\bar{\mathcal{H}}(x)$ by $\pm i \gamma'$:  
\begin{equation}
a(x)=h(x-i\gamma'), \qquad \qquad U(x)=\mathcal{H}(x-i\gamma')=1+a(x),
\end{equation}
 \begin{equation}
\bar{a}(x)=\bar{\mathcal{H}}(x+i\gamma'), \qquad \qquad \bar{U}(x)=\bar{\mathcal{H}}(x+i\gamma')=1+\bar{a}(x),
\end{equation}
where $0<\gamma'<1/2$ is an arbitrary real and fixed parameter. This shift  is necessary because the original functions
$\mathcal{H}(x),\bar{\mathcal{H}}(x)$ have zeroes and poles on the upper and lower half plane respectively.
Due to the ANZ property of  $\hat{T}_{k+1}(x)$ (\ref{86}) and the fact that $\lim_{|x| \to \infty} \frac{d}{dx} \log  \hat{T}_{k+1}(x) = 0$ the following relation holds 
\begin{equation}
0=\int\limits_{-\infty}^{\infty}dx \ \frac{d}{dx} \log  \hat{T}_{k+1}(x-i)\ e^{iq(x-i)}-\int\limits_{-\infty}^{\infty} dx \
\frac{d}{dx} \log  \hat{T}_{k+1}(x+i)\ e^{iq(x+i)}. \label{89}
\end{equation}
From (\ref{78}) and (\ref{79}) one can express $\hat{T}_{k+1}(x \pm i)$ with the auxiliary functions and by substituting these 
expressions  into (\ref{89}) one gets  in  Fourier space 
\begin{eqnarray}
\tilde{dl} \Psi_1 (q>0) &=& \frac{e^q \ \tilde{dl} f(q)}{2 \cosh(q)} 
                                      + \frac{ e^{(2-\gamma')q } }{2 \cosh (q) } \ \tilde{dl} \bar{U} (q) - 
                                                  \frac{ e^{ \gamma' q } }{2 \cosh (q) } \ \tilde{dl} {U} (q) ,  \\
\tilde{dl} \Psi_1 (q<0) &=& 0 ,   \\
\tilde{dl} \Psi_2 (q>0) &=& 0 ,\\
\tilde{dl} \Psi_2 (q<0) &=& - \frac{e^{-q}\  \tilde{dl} f(q)}{2 \cosh(q)}  
                                      - \frac{ e^{-\gamma' q } }{2 \cosh (q) } \ \tilde{dl} \bar{U} (q) + 
                                                  \frac{ e^{-(2-\gamma')q } }{2 \cosh (q) } \ \tilde{dl} {U} (q), 
\end{eqnarray}
where 
\begin{equation}
\tilde{dl} f(q)=e^q \ \tilde{dl} \hat{T}_0^{-(k+2)}(q)-e^{-q} \ \tilde{dl} \hat{T}_0^{+(k+2)}(q).
\end{equation}
We do not need  the explicit form of $\tilde{dl} \hat{T}_0^{\pm (k+2)}(q)$ for our calculations, but we only 
have to know that the following  identities hold
\begin{equation}
e^q \ \tilde{dl} \hat{T}_0^{-(k+2)}(q)=\tilde{dl} \hat{T}_0^{-(k+1)}(q), \qquad \qquad  e^{-q} \ \tilde{dl} \hat{T}_0^{+(k+2)}(q)= \tilde{dl} \hat{T}_0^{+(k+1)}(q).
\end{equation}
After a similar procedure that has been done in the previous section one gets the following nonlinear integral equations in the 
continuum  
\begin{eqnarray}
\log y_1(x) &=& (K*\log Y_2)(x),  \nonumber  \\
\log y_j (x) &=& (K* \log Y_{j-1})+(K*\log Y_{j+1})(x), \qquad  \quad j=2,\dots,k-1   \nonumber  \\
\log y_{k}(x) &=& \mathcal{D}(x)+ (K*\log Y_{k-1})(x)+ (K^{+\gamma'}*\log U)(x) \nonumber  \\
                        &+& (K^{-\gamma'}*\log \bar{U})(x), \label{96}  \\
\log a(x) &=&  (F*\log  U)(x)- (F^{+2(1-\gamma')}*\log \bar{U} )(x) 
                            + (K^{-\gamma'}*\log Y_{k})(x),  \nonumber  \\
\log \bar{a}(x) &=&  (F*\log  \bar{U})(x)- (F^{-2(1-\gamma')}*\log  U )(x)  
                             + (K^{+\gamma'}*\log Y_{k})(x), \nonumber 
\end{eqnarray}
where the \lq\lq source" function and the kernel functions are the same as in the previous section (\ref{60}-\ref{63}).
In the TBA language the complex auxiliary functions $a(x)$ and $\bar{a}(x)$ resum the contributions of 
those Y-system elements whose index is larger than $k$. (See figure \ref{3f}.)

The energy of the continuum model can be expressed by a gauge invariant Y-system element:
\begin{equation}
E_0(l)=E_{bulk}-\frac{m}{4} \int\limits_{-\infty}^{\infty} dx  \cosh \left( \frac{\pi x}{2}\right) \log Y_k(x).
\end{equation}
Equations (\ref{58}) and (\ref{96}) are different descriptions of  the ground
state energy of the current-current perturbation of the $SU(2)_k$ WZW model.
 In this paper one of our main purposes is to describe the finite size 
dependence of the ground state energy of the $O(4)$ NLS model with a closed finite set of nonlinear integral equations. 
According to the light-cone lattice approach \cite{LCS} the $k \rightarrow \infty$ limit of equations (\ref{58}) and (\ref{96}) describes the ground state energy
of the $O(4)$ NLS model. In this limit the equations (\ref{58}) and (\ref{96}) contain an infinite number of unknown functions, but as we 
will see in the next section we are able to construct a finite set of nonlinear integral equations by combining
 (\ref{58}) and (\ref{96})
in the $k \rightarrow \infty$ limit.


\section{Nonlinear integral equations for the $O(4)$ NLS model}

\begin{figure}[htbp]
\begin{center}
\begin{picture}(280,30)(-140,-15)
\put(-170,-10) {\usebox{\kcgn}}
\put(-70,-10) {\usebox{\ofa}}
\put(-135,-15){\parbox{144mm}{\caption{ \label{4f}\protect {\small
a.) Dynkin-diagram associated with the current-current perturbation of the $SU(2)_k$ WZW model, 
 b.) Dynkin-diagram associated with the $O(4)$ NLS model.  }}}}
\end{picture}
\end{center}
\end{figure}
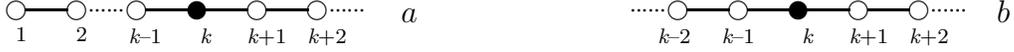

In this section we construct a finite set of nonlinear integral equations from (\ref{58}) and (\ref{96}) in the $k \rightarrow 
\infty$ limit for the ground state of the $O(4)$ NLS model. 
 The complex auxiliary functions (\ref{40}-\ref{41}) and (\ref{76}-\ref{77}) 
 are gauge dependent and have a finite continuum limit only in an appropriate gauge, but the Y-system 
elements (\ref{22}-\ref{23}) are gauge invariant and  have finite continuum limit. One can see from (\ref{42}) and (\ref{82})
 that some gauge 
invariant Y-system elements can be expressed by these non-gauge invariant auxiliary functions:
\begin{equation}
U_0(x+i \gamma) \bar{U}_0(x-i \gamma)=Y_k(x), \label{98}
\end{equation} 
\begin{equation}
U(x+i \gamma') \bar{U}(x-i \gamma')=Y_{k+1}(x).
\end{equation}
If  we choose the Y-system elements (\ref{22}-\ref{23}) ($j=1,2,\dots$) as auxiliary functions, we get the standard TBA 
equations of the model, which can be encoded in an infinite diagram (figure \ref{4f}a.). 

\begin{figure}[htbp]
\begin{center}
\begin{picture}(280,30)(-140,-15)
\put(-165,-7) {\usebox{\suzen}}
\put(-74,-7) {\usebox{\suzene}}
\put(-130,-19){\parbox{130mm}{\caption{ \label{5f}\protect {\small
Graphical representation of the nonlinear integral equations of the $O(4)$ NLS model }}}}
\end{picture}
\end{center}
\end{figure}
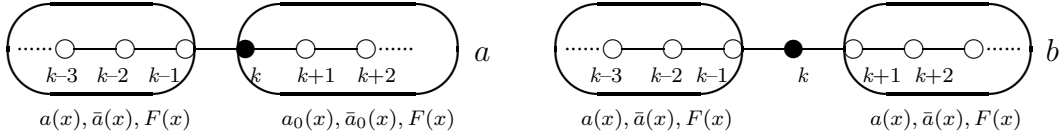

In the $k \rightarrow \infty$ limit the 
diagram will be symmetric with respect to the massive node (figure \ref{4f}b.), so in the $k \rightarrow \infty$ limit $Y_{k-1}(x)=Y_{k+1}(x)$.
So in the $k \rightarrow \infty$ limit the following relations will also be true 
\begin{equation}
U_0(x+i \gamma) \bar{U}_0(x-i \gamma)=Y_k(x),
\end{equation} 
\begin{equation}
U(x+i \gamma') \bar{U}(x-i \gamma')=Y_{k-1}(x). \label{101}
\end{equation}
Substituting (\ref{101}) into the equations for $a_0(x), \bar{a}_0(x)$ in  (\ref{58}) and  (\ref{98}) into the equations for  
$a(x),\bar{a}(x)$ in  (\ref{96}) and deforming  the integration contour appropriately we get a finite set of nonlinear
 integral equations in the $k \rightarrow \infty$ limit, which corresponds to the $O(4)$ NLS model case.
The equations are as follows:
\begin{eqnarray}
\log a_0(x) &=& \mathcal{D}(x-i \gamma)+ (F*\log  U_0)(x)- (F^{+2(1-\gamma)}*\log \bar{U}_0 )(x) \nonumber \\    
                            &+&  (K^{+(\gamma'-\gamma)}*\log U )(x)+ (K^{-(\gamma'+\gamma)}*\log \bar{U} )(x),
 \nonumber  \\
\log \bar{a}_0(x) &=& \mathcal{D}(x+i \gamma)+ (F*\log  \bar{U}_0)(x)- (F^{-2(1-\gamma)}*\log {U}_0 )(x)     \nonumber  \\
                            &+&  (K^{+(\gamma'+\gamma)}*\log U )(x)+ (K^{+(\gamma-\gamma')}*\log \bar{U} )(x),
 \nonumber  \\
\log a(x) &=&  (F*\log  U)(x)- (F^{+2(1-\gamma')}*\log \bar{U} )(x)  \label{102} \\ 
                            &+&  (K^{+(\gamma-\gamma')}*\log U_0 )(x)+  (K^{-(\gamma'+\gamma)}*\log \bar{U}_0 )(x),
 \nonumber  \\
\log \bar{a}(x) &=&  (F*\log \bar{U})(x)- (F^{-2(1-\gamma')}*\log {U} )(x)     \nonumber \\ 
                            &+&  (K^{+(\gamma+\gamma')}*\log U_0 )(x)+ (K^{+(\gamma'-\gamma)}*\log \bar{U}_0 )(x),
 \nonumber  \\
U_0(x)&=&1+a_0(x), \enskip \bar{U}_0(x)=1+\bar{a}_0(x), \enskip U(x)=1+a(x), \enskip
\bar{U}(x)=1+\bar{a}(x). \nonumber 
\end{eqnarray}
 Equations (\ref{102}) are graphically represented in figure \ref{5f}a.
 
The energy expression is the same as (\ref{65}). This is a closed set of equations for four complex unknown functions
\lq\lq $a(x),\bar{a}(x),a_0(x),\bar{a}_0(x)$", but $\bar{a}(x)$  is the complex conjugate of $a(x)$ and  $\bar{a}_0(x)$  
is the complex conjugate of $a_0(x)$, therefore only four real unknown functions describe the model.

From (\ref{96}) one can derive a different set of nonlinear integral equations to describe the finite size dependence of the 
ground  state energy of the $O(4)$ NLS model. Using (\ref{101}) in (\ref{96}) and deforming the integration contour appropriately we get the following equations
 \begin{eqnarray}
\log y_{0}(x) &=& \mathcal{D}(x)+2 (K^{+\gamma'}*\log U)(x)+
                                         2(K^{-\gamma'}*\log \bar{U})(x), \nonumber  \\
\log a(x) &=&  (F*\log  U)(x)- (F^{+2(1-\gamma')}*\log \bar{U} )(x)  \nonumber \\
               &+&  (K^{-\gamma'}*\log Y_{0})(x), \label{103}   \\
\log \bar{a}(x) &=&  (F*\log  \bar{U})(x)- (F^{-2(1-\gamma')}*\log  U )(x) \nonumber \\  
                          &+&  (K^{+\gamma'}*\log Y_{0})(x), \nonumber \\
 U(x)&=&1+a(x), \quad \bar{U}(x)=1+\bar{a}(x), \nonumber 
\end{eqnarray}
where we introduced the notations:
\begin{equation}
y_0(x)=\lim_{k \to \infty} y_k(x), \qquad \qquad Y_0(x)=\lim_{k \to \infty} Y_k(x).
\end{equation}
Equations (\ref{103}) are graphically represented in figure \ref{5f}b. 
In this case the ground state energy of the $O(4)$ NLS model  is of the form
\begin{equation}
E_0(l)=E_{bulk}-\frac{m}{4} \int\limits_{-\infty}^{\infty} dx  \cosh \left( \frac{\pi x}{2}\right) \log Y_0(x).
\end{equation}
 Equations (\ref{103}) contain only three real unknown functions because $y_0(x)$ and $Y_0(x)$ are real and $\bar{a}(x)$ 
 is the complex  conjugate of  $a(x)$. 

Although  equations (\ref{103}) are more convenient for numerical
calculations in the $O(4)$ NLS model than equations (\ref{102}) but, as it will be seen in the next section,  
it is a generalization of (\ref{102}) that describes the finite size dependence of the ground
state energy of the SS-model for finite couplings.


\section{Nonlinear integral equations for the SS-model}

In this section we propose nonlinear integral equations to describe the finite size dependence of the ground state energy 
of the SS-model. In equations (\ref{102})  the kernel function $F(x)$ apart from some trivial 
factor, is nothing but the derivative of the logarithm of the soliton-soliton scattering amplitude at the
 $\beta^2=8 \pi$ point,
\begin{equation}
F(x)=\frac14 \tilde{G}_{\infty} \left(\frac{\pi x}{2}\right), \qquad \tilde{G}_{\infty}(\theta)=\frac{1}{i}
\frac{d}{d \theta} \log  S_{s s} (\theta)\bigg|_{\beta^2 \rightarrow 8 \pi}.
\end{equation} 
 
\begin{figure}[htbp]
\begin{center}
\begin{picture}(280,30)(-140,-15)
\put(-163,-7) {\usebox{\SSddv}}
\put(-74,-7) {\usebox{\SSnlie}}
\put(-130,-15){\parbox{130mm}{\caption{ \label{6f}\protect {\small
Graphical representation of the nonlinear integral equations of the SS-model }}}}
\end{picture}
\end{center}
\end{figure}
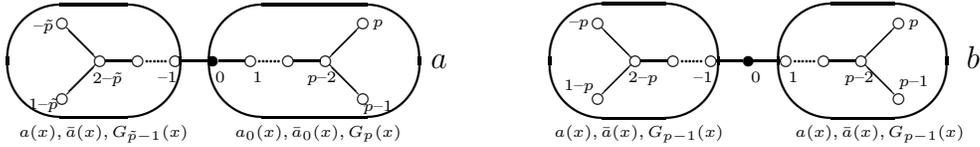

 The scattering matrix of the SS-model is a direct product of two SG S-matrices with different couplings. 
The $O(4)$ NLS model is a special case of the SS-model and its scattering matrix is a direct product of two SG S-matrices
at the $\beta^2 \to 8 \pi$ point. In equations (\ref{102}) we recognize a building block of the S-matrix of the $O(4)$ 
NLS model, as the kernel function $F(x)$. As we mentioned in the introduction the SS-model can be regarded as a two-parameter
deformation of the $O(4)$ NLS model and we assume that the nonlinear integral equations, which describe 
the finite size effects in the SS-model, can also be obtained by a two-parameter deformation of equations (\ref{102}). 
The deformation is given by changing the kernel function $F(x) \to G_p(x)$ with appropriate $p$
value, where $G_p(x)$ is basically the logarithmic derivative of  the soliton-soliton scattering amplitude:
\begin{equation}
G_p(x)=\int\limits_{-\infty}^{\infty} \frac{dq}{2\pi} \ e^{-iqx} \  \frac{\sinh((p-1)q)}{2\cosh(q) \sinh(pq)},
\end{equation}
\begin{equation}
G_p(x)=\frac14 \tilde{G}_{p} \left(\frac{\pi x}{2}\right), \qquad \tilde{G}_{p}(\theta)=\frac{1}{i}
\frac{d}{d \theta} \log S_{ss} (\theta) \bigg|_{\beta^2=\frac{8 \pi p}{p+1}}.
\end{equation}
The conjectured equations are of the form:
\begin{eqnarray}
\log a_0(x) &=& \mathcal{D}(x-i \gamma)+ (G_{p}*\log  U_0)(x)- (G_{p}^{+2(1-\gamma)}*\log \bar{U}_0 )(x) \nonumber \\    
                            &+&  (K^{+(\gamma'-\gamma)}*\log U )(x)+ (K^{-(\gamma'+\gamma)}*\log \bar{U} )(x),
 \nonumber  \\
\log \bar{a}_0(x) &=& \mathcal{D}(x+i \gamma)+ (G_{p}*\log  \bar{U}_0)(x)- (G_{p}^{-2(1-\gamma)}*\log {U}_0 )(x)     \nonumber  \\
                            &+&  (K^{+(\gamma'+\gamma)}*\log U )(x) + (K^{+(\gamma-\gamma')}*\log \bar{U} )(x),
 \nonumber  \\
\log a(x) &=&  (G_{\tilde{p}-1}*\log  U)(x)- (G_{\tilde{p}-1}^{+2(1-\gamma')}*\log \bar{U} )(x) \label{109} \\ 
                            &+&  (K^{+(\gamma-\gamma')}*\log U_0 )(x)+  (K^{-(\gamma'+\gamma)}*\log \bar{U}_0 )(x),
 \nonumber  \\
\log \bar{a}(x) &=&  (G_{\tilde{p}-1}*\log \bar{U})(x)- (G_{\tilde{p}-1}^{-2(1-\gamma')}*\log {U} )(x)     \nonumber \\ 
                            &+&  (K^{+(\gamma+\gamma')}*\log U_0 )(x)+ (K^{+(\gamma'-\gamma)}*\log \bar{U}_0 )(x),
 \nonumber  \\
U_0(x)&=&1+a_0(x), \enskip \bar{U}_0(x)=1+\bar{a}_0(x), \enskip U(x)=1+a(x), \enskip
 \bar{U}(x)=1+\bar{a}(x), \nonumber 
\end{eqnarray}
where $p$ and $\tilde{p}$ are the parameters of action (1) and $0<\gamma \leq 1/2$, $0<\gamma' \leq 1/2$ arbitrary, 
fixed real parameters. Equations (\ref{109}) are graphically represented in figure  \ref{6f}a.

The energy expression is
\begin{equation}
E_0(l)=-\frac{m}{4} \int\limits_{-\infty}^{\infty} dx  \left\{ \cosh \left( \frac{\pi (x-i \gamma)}{2}\right) \log U_0(x) +\cosh \left( \frac{\pi (x+i \gamma)}{2}\right) \log \bar{U}_0(x) \right \}.
\end{equation}
Due to the relations $\bar{a}(x)=a^{*}(x), \ \bar{a}_0(x)=a^{*}(x) $ (* denotes  complex conjugation)
we have only four real unknown functions for all real values of the two couplings $p,\tilde{p}$. The form of  
equations (\ref{109}) is valid for $p \geq 1$ and $\tilde{p} \geq 2$, because in this case the kernel functions of (\ref{109}) have no 
poles in the "main" strip ($0 \leq \mbox{Im} x \leq 1$). 

In contrast to equations (\ref{102}) and (\ref{103}), which were based on a derivation, equations (\ref{109}) are only conjectures.
For this reason in the next section we will give some analytical and numerical evidence showing that equations (\ref{109})
are indeed describing the finite size effects of the ground state energy in the SS-model.


\section{The test of  the equations }

In this section we will make some analytical and numerical tests on our conjectured equations (\ref{109}). First we calculate the UV central charge
of the model using the equations (\ref{109}). Using the standard method of  refs. \cite{KBP,ZamiC} the energy in the conformal limit ($l \rightarrow 0$) 
can be expressed by the dilogarithm functions. The energy in the conformal limit is of the form:
\begin{eqnarray}
E_0(l) \simeq \frac{1}{\pi l} \bigg( L_{+}(a_{+}(\infty))+L_{+}(\bar{a}_{+}(\infty))+L_{+}(a_{0+}(\infty))+L_{+}(\bar{a}_{0+}(\infty)) \\       
-L_{+}(a_{+}(-\infty))-L_{+}(\bar{a}_{+}(-\infty))-L_{+}(a_{0+}(-\infty))-L_{+}(\bar{a}_{0+}(-\infty)) \bigg), \nonumber
\end{eqnarray}
where $L_{+}(z)$ is defined by the integral
\begin{equation}
L_{+}(z)=\frac12 \int\limits_{0}^{z} dx \left( \frac{\log(1+x)}{x}-\frac{x}{1+x} \right), \qquad L_{+}(x)=L \left(\frac{x}{1+x}\right),
\end{equation}
and $L(x)$ is Roger's dilogarithm function. The functions $a_{+}(x),\bar{a}_{+}(x),a_{0+}(x),\bar{a}_{0+}(x)$ denote the 
kink functions corresponding to $a(x),\bar{a}(x),a_{0}(x),\bar{a}_{0}(x)$ respectively. The limits of these kink functions at infinity are as follows:
\begin{equation}
a_{+}(\infty)=\bar{a}_{+}(\infty)=1, \qquad a_{+}(-\infty)=\bar{a}_{+}(-\infty)=0,
\end{equation}
\begin{equation}
a_{0+}(\infty)=\bar{a}_{0+}(\infty)=0, \qquad a_{0+}(-\infty)=\bar{a}_{0+}(-\infty)\rightarrow +\infty.
\end{equation}
From these using the simple identities: 
\begin{equation}
L_{+}(\infty)=L(1)=\frac{\pi^2}{6}, \qquad L_{+}(1)=\frac{\pi^2}{12},
\end{equation}
one gets the ground state energy in the $l \rightarrow 0$ limit
\begin{equation}
E_{0}(l) \simeq -\frac{3 \pi}{6 l}.
\end{equation} 
From this one can easily see that the effective UV central charge of the model is equal to three ($c_{UV}=3$). 
We obtained this value from our conjectured equations (\ref{109}) and it agrees with 
 standard TBA calculations \cite{SS}.   
So we have checked our equations (\ref{109}) analytically in the small $l$ limit in leading order. 

Next we will check our equations in the large $l$ limit analytically in next to leading order.
The equations (\ref{109}) can be solved iteratively in the large $l$ regime. After some easy calculations one gets that the
 ground state energy is of the form
 \begin{equation}
 E_0(l)=E^{(1)}+E^{(2)}+O(e^{-3ml}),  \qquad \qquad E^{(2)}=E^{(2)}_1+E^{(2)}_2, \label{117}
 \end{equation} 
 where
 \begin{equation}
 E^{(1)}=-\frac{4m}{2 \pi} \int\limits_{-\infty}^\infty d\theta \ \cosh(\theta) \ e^{-ml \cosh(\theta)},  
 \end{equation}
 \begin{equation}
 E^{(2)}_1=\frac{16m}{4 \pi} \int\limits_{-\infty}^\infty d\theta \ \cosh(\theta) \ e^{-2ml \cosh(\theta)},  
 \end{equation}
 \begin{equation}
 E^{(2)}_2=-\frac{m}{4 \pi^2} \int\limits_{-\infty}^\infty d\theta \ \cosh(\theta) \ e^{-ml \cosh(\theta)}
 \ \int\limits_{-\infty}^\infty d\theta' \ \Phi(\theta-\theta') \ e^{-ml \cosh(\theta')},
 \end{equation}
where the kernel function $\Phi(\theta)$ is of the form
\begin{equation}
\Phi(\theta)=\int\limits_{-\infty}^\infty d \omega \ \tilde{\Phi}(\omega) \ e^{-i \omega \theta}, \qquad \qquad  
\tilde{\Phi}(\omega)= \tilde{\Phi}_{p}(\omega)+ \tilde{\Phi}_{\tilde{p}}(\omega), 
\end{equation}
where
\begin{equation}
\tilde{\Phi}_{p}(\omega)=4-8 \ \frac{\sinh^2 \left( \frac{\pi \omega}{2}  \right) \ \sinh \left(\frac{(p-1)\pi \omega}{2} \right) }
{\cosh \left(\frac{\pi \omega}{2} \right) \ \sinh \left( \frac{p \pi \omega}{2} \right) }. \label{122}
\end{equation}
The same results can be obtained from the TBA equations of the SS-model with integer parameters $p,\tilde{p}$ using the method of  \cite{viri}. One can recognize that $\Phi(\theta)$ is nothing but $1/i$ times the trace of the
logarithmic derivative of the two-body S-matrix of the SS-model:   
\begin{equation}
\Phi(\theta)=\frac{1}{i} \ \frac{d}{d \theta} \ \mbox{Tr}_2 \log S(\theta),
\qquad S(\theta)=-(S_{p} \otimes S_{\tilde{p}})(\theta). \label{123}
\end{equation}
It is well-known that the ground state energy in finite volume is related to the free energy density of the system 
at temperature $T$ by $f(T)=E_0(1/T)T$. The free energy density can be expressed as a series in the low
temperature limit (virial expansion) and the coefficients of the series are completely determined by S-matrix data \cite{Dashen}. 
If we transform the next to leading order large $l$ expansion of the ground state energy (\ref{117}-\ref{122}) into the next to leading
 order low temperature expansion of the free energy, we get perfect agreement with the prediction of the virial
expansion \cite{Dashen}.  

So far we have made some analytical tests on our proposed equations (\ref{109}) in the small $l$ and large $l$ regimes.
In the intermediate
regime we solved numerically the equations (\ref{109}) for some integer values of $p$ and $\tilde{p}$, and we have solved the corresponding
TBA equations, too. In every  case  we examined perfect agreement  between the two sets of numerical data 
was found.  
 

\section{Special cases and forms}

In this section we consider first the special case $p=\tilde{p}$. In this case instead of (\ref{109}) it is better to use 
the following modified version of  (\ref{103})
 \begin{eqnarray}
\log y_{0}(x) &=& \mathcal{D}(x)+2 (K^{+\gamma'}*\log U)(x)+
                                         2(K^{-\gamma'}*\log \bar{U})(x), \nonumber  \\
\log a(x) &=&  (G_{p-1}*\log  U)(x)- (G_{p-1}^{+2(1-\gamma')}*\log \bar{U} )(x) \nonumber  \\
               &+&  (K^{-\gamma'}*\log Y_{0})(x),   \label{124} \\
\log \bar{a}(x) &=&  (G_{p-1}*\log  \bar{U})(x)- (G_{p-1}^{-2(1-\gamma')}*\log  U )(x)  
                             + (K^{+\gamma'}*\log Y_{0})(x), \nonumber \\
 U(x)&=&1+a(x), \quad \bar{U}(x)=1+\bar{a}(x)\quad Y_0(x)=1+y_0(x). \nonumber 
\end{eqnarray} 
The graphical notation of eqs. (\ref{124}) can be seen in figure \ref{6f}b.
The energy is of the form
\begin{equation}
E_0(l)=-\frac{m}{4} \int\limits_{-\infty}^{\infty} dx  \cosh \left( \frac{\pi x}{2}\right) \log Y_0(x).
\end{equation}
These equations are valid for the $p \geq 2$ case and they are better for numerical calculations than (\ref{109}) 
because the number of unknown functions is less than in (\ref{109}). 

 Next we consider the special form of equations (\ref{109}), which corresponds to the $N=2$ supersymmetric SG model.
 This model is a special case of the SS-model. The scattering matrix of the model is of the form \cite{kobi}
 \begin{equation}
 S_{SG}^{N=2}(\theta)\bigg|_{\tilde{\beta}^2=8 \pi p}=-(S_2 \otimes S_p)(\theta),
 \end{equation} 
where $\tilde{\beta}^2$ is the coupling constant of the SUSY-SG model. We can see that the S-matrix of this theory is
a special case of the S-matrix (\ref{3}) of the SS-model. Applying equations (\ref{109}) for this case and taking $\gamma' \rightarrow 0$ we obtain
\begin{eqnarray}
\log a_0(x) &=& \mathcal{D}(x-i \gamma)+ (G_{p}*\log  U_0)(x)- (G_{p}^{+2(1-\gamma)}*\log \bar{U}_0 )(x) \nonumber \\    
                            &+&  (K^{-\gamma}*\log U )(x)+  (K^{-\gamma}*\log \bar{U} )(x),
 \nonumber  \\
\log \bar{a}_0(x) &=& \mathcal{D}(x+i \gamma)+ (G_{p}*\log  \bar{U}_0)(x)- (G_{p}^{-2(1-\gamma)}*\log {U}_0 )(x)     \nonumber  \\
                            &+&  (K^{+\gamma}*\log U )(x) + (K^{+\gamma}*\log \bar{U} )(x),  \\
\log a(x) &=&   (K^{+\gamma}*\log U_0 )(x) + (K^{-\gamma}*\log \bar{U}_0 )(x), \nonumber  \\
\log \bar{a}(x) &=&  (K^{+\gamma}*\log U_0 )(x)+ (K^{-\gamma}*\log \bar{U}_0 )(x). \nonumber  \\
U_0(x)&=&1+a_0(x), \enskip \bar{U}_0(x)=1+\bar{a}_0(x), \enskip U(x)=1+a(x), \enskip 
\bar{U}(x)=1+\bar{a}(x). \nonumber 
\end{eqnarray}  
It can be seen that in this case $a(x)=\bar{a}(x) \in \mathbb{R}$, and we get the same equations as the ones proposed by
Dunning \cite{dun} for the $N=2$ SUSY-SG model.


\section{Summary and Conclusions}

From the isotropic higher spin six-vertex model we derived two different sets of nonlinear integral 
equations  describing the finite size dependence of the ground state energy of   the 
current-current perturbation of the $SU(2)_k$ WZW model. In the $k \to \infty$ limit we glued the
two set of equations together and obtained a finite set of nonlinear integral equations for the $O(4)$ NLS
model. Starting from these equations we proposed a new set of nonlinear integral equations for describing
finite size effects of the SS-model. We made analytical and numerical tests on our conjectured equations and
all the results of these tests made us confident that our conjectured equations are correct.  The advantage of our 
equations is that the number of unknown functions is minimal, and the equations can be defined for all real values 
of the  coupling constants. 

It would be interesting to generalize these equations for describing (all) excited states energies
of the model.

\vspace{1cm}
{\tt Acknowledgements}

\noindent 
I would like to thank J\'anos Balog for useful discussions. 
This investigation was supported in part by the 
Hungarian National Science Fund OTKA (under T034299
and T043159).

\appendix

\section{Appendix }

In this appendix we summarize our conventions of Fourier transformation. Let $f(x)$ be a bounded continuous function defined
on the whole real axis. Its Fourier transform is defined as:
\begin{equation}
\tilde{f}(q)=\int\limits_{-\infty}^{\infty} dx \ e^{iqx} \ f(x)
\end{equation}  
and the inverse Fourier transform is
\begin{equation}
f(x)=\int\limits_{-\infty}^{\infty} \frac{dq}{2 \pi} \ e^{-iqx} \ \tilde{f}(q).
\end{equation}
The convolution of two functions is defined as follows:
\begin{equation}
(f*g)(x)=\int\limits_{-\infty}^{\infty} dy \ f(x-y)g(y).
\end{equation}
The Fourier transform of such a convolution is the product of the Fourier transforms of the two functions.

\end{document}